\documentclass[12pt,preprint]{aastex}

\advance \voffset by 1.00cm\relax
\def\msun{{\,M_{\odot}}}

\def\mpy{{\,M_{\odot}\,{\rm yr}^{-1}}}
\def\kms{{\,{\rm km}\,{\rm s}^{-1}}}

\newcommand{\Fewbody}{{\em Fewbody}}



\begin{document}

\title{The Total Merger Rate of Compact Object Binaries In The Local Universe}

\author{Aleksander Sadowski\altaffilmark{1}, Krzysztof
Belczynski\altaffilmark{2,3}, 
        Tomasz Bulik\altaffilmark{1,4}, Natalia Ivanova\altaffilmark{5}, 
	Frederic A.\ Rasio\altaffilmark{6}, Richard O'Shaughnessy\altaffilmark{7}}

\email{as@camk.edu.pl, kbelczyn@nmsu.edu, tb@astrouw.edu.pl, nata@cita.utoronto.ca, rasio@northwestern.edu, oshaughn@gravity.psu.edu}

     \altaffiltext{1}{Nicolaus Copernicus Astronomical Center,
            Bartycka 18, 00-716 Warszawa, Poland}
     \altaffiltext{2}{Los Alamos National Laboratory (Oppenheimer Fellow), Los Alamos, NM, USA}
     \altaffiltext{3}{New Mexico State University, Las Cruces, NM, USA}
     \altaffiltext{4}{Astronomical Observatory, Warsaw University,
            Al.\ Ujazdowskie 4, 00-478, Warsaw, Poland}
     \altaffiltext{5}{Canadian Institute for Theoretical Astrophysics,
            University of Toronto, 60 St. George, Toronto, ON M5S 3H8, Canada}
     \altaffiltext{6}{Northwestern University, Dept of Physics and Astronomy, 2145 Sheridan Rd, Evanston, IL 60208}
     \altaffiltext{7}{Pennsylvania State University, Department of Physics, 104 Davey Laboratory, University Park, PA 16802}

\begin{abstract} 
Using a population synthesis approach, we compute the total merger rate in the local Universe for 
double neutron stars, double black holes, and black hole -- neutron star binaries. 
These compact binaries are the prime source candidates for gravitational-wave
detection by LIGO and VIRGO. We account for mergers originating {\em both\/} from
field populations and from dense stellar clusters, where dynamical
interactions can significantly enhance the production of double compact objects.
For both populations we use the same treatment of stellar evolution. 
Our results indicate that the merger rates of double neutron 
stars and black hole -- neutron star binaries are strongly dominated by field
populations, while merging black hole binaries are formed much more effectively 
in dense stellar clusters. The overall merger rate of double compact objects 
depends sensitively on the (largely unknown) initial mass fraction
contained in dense clusters ($f_{\rm cl}$). For  $f_{\rm cl} \lesssim 0.0001$,
the Advanced LIGO detection rate will be dominated by field populations of
double neutron star mergers, with a small but significant number of detections
$\sim 20$ yr$^{-1}$. However for a higher mass fraction in clusters,
$f_{\rm cl} \gtrsim 0.001$, the detection rate will be dominated by numerous 
mergers of double black holes originating from dense clusters, and it will be 
considerably higher, $\sim 25 - 300$ yr$^{-1}$.
In addition, we show that, once mergers of double black holes
are detected, it is easy to differentiate between systems formed in the
field and in dense clusters, since the chirp mass distributions are
strikingly different. If significant field populations of double black
hole mergers are detected, this will also place very strong constraints on
common envelope evolution in massive binaries. Finally, we point out that
there may exist a population of merging black hole binaries in  
intergalactic space. 
\end{abstract}


\section{INTRODUCTION}

Gravitational wave astronomy is entering a new era:
LIGO (Abramovici et al.\ 1992) has now taken a full year of data at its design 
sensitivity; VIRGO (Bradaschia et al.\ 1990) is nearing completion. Other detectors
like GEO or TAMA are also on track. There are well defined plans
for improving the LIGO and VIRGO detectors so that, within the next few years, their
sensitivity  will increase by a factor of up to 30. 
The list of potential sources for these high-frequency detectors is long and includes
supernova explosions, neutron star oscillations, and persistent
radiation from rapidly rotating, nonaxisymmetric neutron stars.
However, compact object binaries remain the most promising sources.
It has been known for some time that their formation 
may take place in two very different environments:
in the galactic field, where their progenitors are massive binaries that evolve in isolation,
and in dense star clusters, where they can form at high rates through dynamical interactions.

The properties of the populations of double compact object binaries
have been previously investigated using both observational and theoretical approaches.
In the typical observational approach, the known 
compact objects binaries, i.e., radio pulsars in double neutron star (NS-NS) systems, were analyzed
in detail. Based on their observed properties and the radio selection 
effects the properties of the entire population can be reconstructed and the 
expected merger rate can be calculated (Kim et al.\ 2005).
This approach, however, cannot be extended to black hole -- neutron star (BH-NS) binaries and 
double black hole  (BH-BH) binaries since these systems have never been observed. 
Moreover, we know only one merging NS-NS binary in a globular cluster, 
and so this method has little power for the population
originating in clusters.  The second, theoretical, approach 
is based on numerical simulations of stellar evolution in field populations, combined with
 dynamical simulations of star clusters. For field populations some recent examples of theoretical
studies include those by Belczynski, Kalogera \& Bulik
(2002, hereinafter BKB02), Voss \& Tauris (2003), Pfahl et al.\ (2005), Dewi et al.\ (2006) and
Belczynski et al.\ (2007a).

The importance of globular cluster evolution for double compact object
formation has long been suspected. The fate of BH populations in globular
clusters was studied by a number of groups (e.g., Sigurdsson \& Hernquist 1993; 
Kulkarni, Hut \& McMillan 1993;
Portegies Zwart \& McMillan 2000; Merritt et al.\ 2004). However, previous
studies have employed only very simplified treatments of stellar
evolution for single stars and binaries in clusters, focusing instead mostly on 
dynamical interactions.   
It was pointed out (Phinney et al.\ 1991; and more recently Grindlay,
Portegies Zwart \& McMillan 2006) that the formation of NS-NS and BH-NS
binaries in clusters is not very efficient and the contribution of clusters
to their total merger rate is rather small ($\sim 10-30 \%$). 
Although the predicted formation rate of merging double NS has a strong density
dependence (Ivanova et al.\ 2007), even for dense clusters like
like 47~Tuc the NS contribution to total cluster merger rates should be overwhelmed
by double BH mergers. Such systems are expected to be formed via dynamical
interactions in cluster cores (Gultekin, Miller \& Hamilton 2004; O'Leary et al.\ 2006, 2007).
O'Leary et al.\ (2006, 2007) used realistic initial conditions for their cluster 
simulations (which included the stellar evolution of massive stars and binaries in the first few 
Myr of the cluster life) and assumed immediate creation of a BH subcluster. Their calculations
did not consider further stellar or binary evolution and neglected the effects of lower-mass stars on BH populations. Their results provide the most up-to-date estimates of BH-BH merger rates from 
globular clusters and the corresponding LIGO detection rates assuming rapid BH segregation
 into an isolated subcluster. 
Mackey et al.\ (2007) investigated the effects of BHs on the structural evolution
of globular clusters. They used a realistic initial mass function for single stars and observed rapid
mass segregation of the BHs into the cluster core (as expected from the Spitzer instability; see 
Watters et al.\ 2000 and references therein).
However, this study was based on direct $N$-body simulations that did not include {\em any\/} primordial
binaries, even though binaries could affect the mass segregation and subsequent dynamics very 
significantly.

In this paper we take the next step and calculate the evolution of representative 
star clusters from the onset of star formation taking into account the large binary fraction for
massive stars. We assume that inelastic interactions of hard binaries
in the cluster core are effective enough to prevent BHs from completely separating from the rest
of the cluster, and we therefore treat the BHs as always well mixed and in thermal equilibrium with other stars
in the cluster core. This simplifying assumption is opposite, and complementary, to the one adopted 
in the models of O'Leary et al. (2006, 2007).
Their results and the merger rates calculated in this work give, respectively, lower and upper bounds 
on the LIGO detection rates. 

In our new models we include both stellar dynamics and full stellar evolution for single 
and binary stars to predict the merger rate of double compact objects. All 
stellar populations are evolved and allowed to interact through an entire 
cluster lifetime ($\sim 13$ Gyr).  
Then we combine our estimates for star clusters and the merger rates calculated in O'Leary
 et al.\ (2006) with the most recent population synthesis field calculations  (Belczynski et al.\ 2007a) 
to deduce the total merger rate. In Section~2 
we present a description of the stellar binary evolution treatment used and we introduce 
our model for globular clusters. In Section~3 we present our results and in Section~4 
a brief discussion.

\section{MODEL DESCRIPTION}
\label{assumptions}

For field merger rates we use the recent calculations of Belczynski et al.\ (2007a) 
obtained with the revised population synthesis code {\tt StarTrack}
(see \ref{startrack}). To estimate the merger rates in dense (globular) 
clusters we employ the dynamical code developed by Ivanova et al.\ (2005; see
also \ref{dynamics}) to calculate a typical cluster model (see 
\ref{standmodel}).

\subsection{Stellar evolution}
\label{startrack}

Our investigation is based on a population synthesis approach. We use the
{\tt StarTrack} population synthesis code (BKB02), 
which has been revised and improved significantly 
over the past few years (Belczynski et al.\ 2007b). 

All stars are evolved based on the metallicity- and wind-mass-loss-dependent 
models of Hurley, Pols \& Tout (2002), with a few improvements described in 
BKB02. The main code parameters we use correspond to the standard model presented 
in Belczynski et al. (2007b). Each star, either single or a binary component, is 
placed initially on 
the zero-age main sequence (ZAMS) and then evolved  through a sequence of distinct 
phases: main sequence (MS), Hertzsprung Gap (HG), red giant branch (RG), core He 
burning (CHeB), asymptotic giant branch (AGB); if a star gets stripped of its 
H-rich envelope, either through wind mass loss or Roche lobe overflow (RLOF) it 
becomes a naked helium star (He). The nuclear evolution leads ultimately to the 
formation of a compact object. Depending on the pre-collapse properties of a
star and its composition this may be a white dwarf (WD), a neutron star (NS) or 
a black hole (BH). 

We model the following processes, which can alter the binary orbit and subsequent 
evolution of the binary components: tidal interactions, magnetic braking, 
emission of gravitational radiation, and angular momentum changes due to mass loss. 
Binary components may interact through mass transfer and accretion phases. We take 
into account various modes of mass transfer: wind accretion and Roche lobe
overflow (RLOF); conservative and non-conservative; stable or dynamically unstable 
(leading to common-envelope evolution). The mass transfer rates are calculated 
from the specific binary configurations and physical properties (masses, evolutionary 
stages, etc.) of the stars involved. Binary components may loose or gain
mass, while the binary orbit may either expand or tighten in response, depending
on the particular details of the mass transfer.

Several modifications of the code relevant for double compact object formation  were 
recently introduced: we employ a self-consistent compact object mass calculation 
(Belczynski et al.\ 2007b) and spin evolution of BHs in compact binaries is followed 
(Belczynski et al.\ 2007c).
Additionally, we have taken into account stellar structure of stars crossing
Hertzsprung gap and we have pointed out that if such a star initiates common
envelope phase it most likely leads to a merger dramatically decreasing
formation rates of close double black hole binaries (for more details see 
Belczynski et al.\ 2007b).

\subsection{Dynamical interactions}
\label{dynamics}

A full description of our Monte Carlo procedure to treat the dynamics of dense
star clusters is given in Ivanova et al.\ (2005). The most recent updates to the code
are described in Ivanova et al.\ (2006, 2007).
The method incorporates a 
full treatment of binary and single star evolution (as described in \ref{startrack}) and 
allows the dynamical evolution of very large ($N\sim 10^6$) 
systems with arbitrary initial binary fractions (e.g., $\sim 50\%$) to be computed in
$\sim$ 1 month on a modern workstation with 4 processors.

Our modeling of the cluster dynamics is based on several assumptions.  
We assume that the core number density, $n_{\rm c}$, and one-dimensional
velocity dispersion, $\sigma$, remain constant throughout the
evolution (the cluster remains in a constant state of thermal equilibrium for its
entire evolution). Typically, the density and "temperature" profiles of a cluster 
do not change much as long as there are enough binaries remaining to provide support
against gravothermal contraction. This state can last for many tens of half-mass
relaxation times ($\sim~10^9 yr$ for most Galactic globular clusters)
as has been shown in many studies (e.g., Fregeau et al.\ 2003;
Giersz \& Spurzem 2003; Hurley \& Shara 2003).
It is not clear that the cluster BH population can in fact remain in thermal equilibrium
with other stars in the presence of many binaries. However, as mentioned earlier, we make 
this simplifying assumption here.
The core number density and velocity dispersion are input parameters used to calculate
dynamical interaction rates in the cluster core (see below). While all
globular clusters have $\sigma_{1D} \sim 10\,{\rm km}\,{\rm s}^{-1}$, the
core density can vary by several orders of magnitude reaching $\sim 10^6\,{\rm pc}^{-3}$
or even higher in some of
the densest clusters (e.g., Harris 1996).

The escape velocity from the cluster core can be estimated from
observations as $v_{\rm e} = 2.5\,\sigma_{3D}$ (Webbink 1985), where
$\sigma_{3D}$ is the three-dimensional core velocity dispersion.
Following a dynamical interaction or a supernova explosion, any object
(single or a binary) that has acquired a recoil speed exceeding $v_{\rm e}$ is 
moved outside the cluster where it continues its evolution in isolation.

  For computing interactions in the core, the velocities of
all objects are assumed to be distributed according to a down-scaled 
Maxwellian (King 1965), with 
$f(v) = v^2/\sigma(m)^2 (\exp(-1.5 v^2/\sigma(m)^2) 
- \exp(-1.5 v_e^2/\sigma(m)^2))$ for $v<v_e$ and $f(v)=0$ for $v>v_e$
with parameters
$\sigma(m)= (\langle m \rangle/m)^{1/2}\sigma_{3D}$ (assuming energy
equipartition in the core) and $v_{\rm e}$. In addition, we use
$\sigma$ to impose a cut-off for soft binaries entering the core.  Any
binary with maximum orbital speed $< 0.1 \sigma_{3D}$ is immediately
broken into two single stars (Hills 1990).

To model mass segregation in our simulations, we assume that the
probability for an object of mass $m$ to enter the core after a time
$t_s$ follows a Poisson distribution,
$p(t_s)=(1/t_{sc})\exp(-t_s/t_{sc})$, where the characteristic
mass-segregation timescale is given by $t_{sc}=10\ C\ \left(\langle m
\rangle/m\right) t_{\rm rh}$, Fregeau et al. 2002). Here $t_{\rm rh}$
is the half-mass relaxation time, which we assume to be constant for a
given cluster.  We fix $C\ \langle m \rangle=3 M_\odot$, as this value
gives, in our model, the best fit for the ratio of core mass to total
cluster mass in 47~Tuc.

All objects are allowed to interact dynamically after they have 
entered the cluster core. We use a Monte-Carlo prescription
to decide which pair of objects actually have an interaction
during each time step.
We consider separately binary--binary and binary--single interactions, 
as well as single--single encounters (tidal captures and collisions).
Tidal captures are treated using the approach described by Portegies Zwart 
\& Meinen (1993).
If the pericenter distance is less than twice the sum of the stellar radii,
the encounter is treated as a physical collision and assumed to lead to a merger.
Each dynamical interaction involving a binary is calculated using
\Fewbody, a numerical toolkit for simulating small-$N$ gravitational
dynamics that is particularly suited to performing 3-body and 4-body 
integrations (Fregeau et al.\ 2004).

\subsection{Standard Cluster Model}
\label{standmodel}

We draw initial masses of single stars and primary stars in binaries using a
three-component Kroupa (2002) power-law initial mass function with the following 
exponents: $\alpha_{1}=-0.3$ for $m/\msun<0.08$, $\alpha_{2}=-1.3$ for $0.08\leq 
m/\msun<0.5$ and $\alpha_{3}=-2.3$ for stars more massive than $0.5\msun$. 
We assume that the minimum object mass is $0.05\msun$\footnote{We allow some
objects to reach a mass below the hydrogen-burning limit, $0.08\msun$, to avoid
skewing the binary mass-ratio distribution.}. The stars are allowed 
to reach masses as high as $150\msun$. All stars are formed in a single starburst 
at the beginning of our simulation. The initial binary mass ratio is assumed to be 
flat between 0 and 1. Binary periods are drawn from uniform distribution in $\log_{10}P$ 
in the range from $0.1$ up to $10^7$ days. However, binaries with either
components filling its Roche lobe are immediately rejected. We assume thermal 
distribution for eccentricities ($\sim 2e$). 
We set the initial binary fraction to $f_{\rm bi}=50\%$ (the number of single stars 
is equal to the number of binaries). Stars are evolved with low metallicity
of $Z=0.001$ appropriate for globular cluster environment (Harris 1996) using 
standard wind mass loss rates (Hurley et al. 2002) modified by 
Belczynski (2007b) to include winds from low and intermediate mass main
sequence stars. We apply the energy prescription with $\alpha_{CE}\times\lambda = 1.0$ 
for common envelope evolution (Webbink 1984). We assume maximal NS mass to be 2.5 $\msun$.
We adopt natal kicks from Hobbs et
al.\ (2005) for NSs, decreased natal kicks for fall-back BHs and 
no kicks for direct collapse BHs (details can be found in Belczynski et al.\ 2007b). The
merger of two BHs forms a single BH that receives a recoil kick due
to anisotropic emission of gravitational radiation during the merger.  We use
results of Sopuerta et al.\ (2007; their best estimate model) to estimate the
magnitude of this kick. 

Cluster properties are defined by the number of stars, core density, half-mass 
relaxation time, one-dimensional velocity dispersion and initial core mass fraction. 
In our model we adopt values  typical for a dense globular cluster in our Galaxy. 
We evolve $10^6$ stars (both single and binary stars) with binary fraction
of $50\%$. The total initial mass of the 
cluster is $4.82 \times 10^5\msun$. 
This value is higher than the average mass of current Milky Way
globular clusters (Gnedin \& Ostriker 1996). However, one has to remember 
that globular clusters lose a significant fraction of their mass 
during their evolution due to two-body relaxations, gravitational shocks and stellar evolution.
The mass loss can be as high as 50\% in a Hubble time (Fall \& Zhang 2001) and therefore
our cluster can be considered typical after a few Gyr of its evolution.

We set the core density to $10^5\,{\rm pc}^{-3}$. 
This choice of density sets the importance of dynamical interactions. 
The chosen value corresponds to a fairly dense cluster like 47~Tuc.
Additionally, we have calculated a cluster model with mass five times lower than the mass of
our standard cluster and the merger rate of BH-BH binaries per unit mass did
not changed significantly (see below).  
The half-mass relaxation time, which determines the rate of mass segregation, is 
assumed to be $10^9\,$yr. The cluster escape velocity is determined by one-dimensional 
velocity dispersion $\sigma_{1D}$. We set $\sigma_{1D}=11.54\kms$ which results in 
$v_{e}=50\kms$. We also assume that initially 10\% of the cluster mass is located in the 
core.

Our calculations with a large fraction of binaries and comprehensive stellar evolution 
for a very large number of stars in a cluster are computationally expensive. Therefore, we 
limited this study to 5 independent runs. The obtained merger rates were then averaged.
Our results presented here are based on these mean values. To assess the influence of initial 
cluster mass on the
cluster merger rate we performed a calculation for a small cluster (initially $2\times10^5$ stars,
corresponding to a mass $\sim 10^5\msun$) and we found that the merger rates scale down
roughly proportionally to mass. We were not able to perform similar calculations for
more massive clusters. But one can expect that for massive clusters the rates
(again per unit mass) are not lower than for our standard cluster.  Therefore, our results can be 
considered as reasonably robust estimates for the contribution of clusters to the overall merger 
rates.

\section{RESULTS}
\label{results}

\subsection{Cluster BH-BH formation}
\label{channels}

All types of double compact object mergers can be produced in a star cluster.
Nevertheless, the most massive objects are most likely participants in dynamical interactions, which
leads to the formation of massive tight binaries (Fregeau et al.\ 2004). Therefore, 
BH-BH mergers are expected to strongly dominate the overall merger rate from globular 
clusters (over lighter BH-NS and NS-NS systems).

\begin{figure}
\epsscale{.95}
\plotone{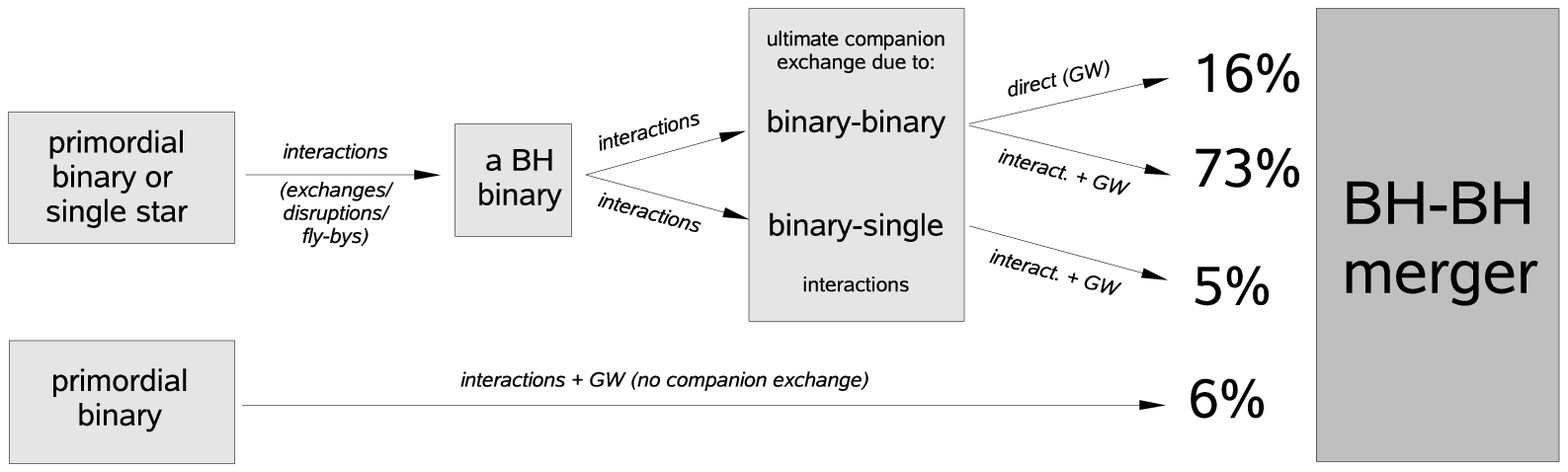}
\caption{
Channels leading to BH-BH mergers in clusters with their contributions. 'Interactions' means any kind of 
dynamical interactions with the exception of the primordial BH binary channel where no companion 
exchange takes place; 'GW' means binary tightening due to gravitational wave emission. For details 
see \S\ref{channels}.
}
\label{f.channels}
\end{figure}

In Figure~\ref{f.channels} we show various evolutionary scenarios leading to
BH-BH binary mergers.
These mergers can be formed either from primordial massive binaries (6\%) or
through exchange interactions between single and/or binary stars (94\%). 
The densities of typical globular cluster cores are high enough to
ensure that all BHs interact frequently with other objects. Most of these
encounters involve low-mass MS stars (these stars are so numerous that
they dominate small cross-section interactions). However, once a close BH-BH binary is formed
no star is able, in general, to disrupt it. Dynamical encounters strongly influence
the properties of young BH populations in clusters.
The most common interactions, leading to the formation of close BH-BH
systems, include encounters of two binaries. Such a binary--binary interaction often leads
to a companion exchange (low-mass MS star is replaced by a massive BH).
Subsequent interactions (fly-by's) further hardens these systems.
At some point the binary becomes tight enough to merge in a short time
through angular momentum loss by gravitational radiation.
Although all stellar mass BHs are formed very early in the evolution
of a cluster (first $\sim 10-20 {\rm Myr}$), the formation of BH-BH binaries may be
extended for a long time due to the ongoing dynamical interactions which create
double BH binaries through various evolutionary channels.

It is expected that very few coalescing (in a Hubble time) BH-BH 
binaries can be formed without dynamical interactions in the field (Belczynski et al.\
2007a). However, the field population contains a large number of BH-BH
binaries that are too wide to coalesce in a Hubble time. We find that, even if
dynamical interactions are included, globular cluster environments form
initially similar populations of BH-BH binaries (mostly wide). However, once the
dynamical interactions become more effective in the cluster core, many of 
these binaries are disrupted (providing single BHs to the cluster population)
and a small but  significant fraction is hardened by interactions (fly-by's) 
with other stars (mostly binaries) and evolves to form close BH-BH systems
(here about $6\%$ of the entire cluster population; see Fig.~\ref{f.channels}).  

Let us follow an example of a typical evolution that leads to a BH-BH merger. The 
primordial massive star ($M_{\rm ZAMS} \sim 25\msun$) either is formed in or
finds its way to the cluster core in a short time. Its evolution leads to a supernova 
explosion leaving a BH of mass $M_{\rm BH,1} \sim 20\msun$.
Note that our models are calculated for low metallicity. That leads to 
reduced mass loss and potential direct BH formation even at masses as low as
25 Msun. Detailed discussion of BH mass spectrum at low metallicity was
presented by Belczynski et al. (2004).
The BH soon becomes a component of a binary as 
a result of a binary-single exchange. After 40 Myr the BH exchanges its
light binary companion for a MS star of mass $\sim 7\msun$. The new companion 
explodes as a supernova after 87 Myr, disrupting the binary. So far the BH and its binaries 
have undergone $\sim 20$ significant binary--single and 4 binary--binary fly-by interactions in addition
to the exchanges. Through the supernova explosion the BH obtains a significant recoil 
velocity and is ejected from the cluster core into the halo. It takes another 400 Myr for mass 
segregation to bring it back to the core. At 5 Gyr the BH acquires again a binary companion. 
The relatively light 
companion ($M=0.87\msun$ MS star) is soon exchanged  for 
another massive BH ($M_{\rm BH,2}=21 \msun$; this is the last exchange in the history). 
In the next few Gyr the BH-BH binary is subject
to $\sim 80$ binary--single and $\sim 20$ binary--binary fly-by interactions that make the 
binary orbit harden significantly. After almost 8 Gyr of cluster evolution this BH-BH binary 
finally merges through gravitational-wave emission. With a mass ratio of
merging BHs near unity the merger product (single BH) will likely stay in the cluster
(i.e., it does not receive a large recoil kick; see \S\,3.2) and is therefore subject
to further interactions.

\subsection{Properties of coalescing BH-BH binaries}

Figure~\ref{f.merg} presents the number of double BH mergers as a function
of cluster age in our standard cluster model. 
The solid line shows the numbers of all mergers while the dashed line
represents mergers originating from primordial binaries. We show number of
mergers per 1 Gyr time bin, so the plot can also be read as a merger rate 
in Gyr$^{-1}$. One can see that the overall rate is oscillating around the 
mean value of $\sim 2.5$ Gyr$^{-1}$. The fluctuations are large but in general 
the merger rate distribution is statistically consistent with the constant rate. 
Moreover, the globular clusters star formation was assumed to be a
$\delta$-function, but if it was spread over some time the merger rate would
be even more constant over a long period of time ($\sim$ Hubble time).

The primordial BH-BH mergers occur in binaries that survived supernova explosions 
of both components. The BH-BH formation (from primordial stars) happens in the first 
$\lesssim 20$ Myr of the cluster evolution. Thus, one could expect that further tightening due 
to dynamical encounters and gravitational-wave emission will result in a merger in 
a relatively short time. In fact, the rate of primordial BH mergers (dashed line in Fig.~\ref{f.merg})
falls with the age of the cluster, and in particular after
$\sim$ 5 Gyr of the cluster evolution there is no or very little mergers from
primordial binaries.

Few-body interactions favor creation of massive binaries. Therefore, BH-BH binaries 
that are formed through these dynamical interactions contain most massive BHs. Mergers 
of such systems are characterized by very high chirp masses. Their distribution is shown in
Figure~\ref{f.mchirp}. The BH-BH mergers are found with chirp masses of $\sim 10-30 
\msun$ with the peak of a distribution between $18$ and $20\msun$. These values are 
significantly higher than expected for the Galactic field BH-BH population 
($\sim 7 \msun$; Belczynski et al 2007a). The reason behind a significant change of the chirp mass distribution
(from field to cluster population) lies not only in changes of binary 
mass-ratio distribution caused by dynamical interactions but also in the 
fact that most massive single black holes are captured into binaries to 
form massive double black hole binaries. However, it also needs to be noted 
that black holes in clusters are on average much more massive than in our 
field simulations as we assume low metallicity of old globular clusters. In 
particular, for solar metallicity in field populations we form single black 
holes up to $\sim 10 \msun$, while for typical globular cluster metallicity 
single black holes reach $\sim 20 \msun$ (e.g., Belczynski et al. 2004).

In our models the recoil kick imparted to the product of double BH mergers is calculated as in 
Sopuerta et al.\ (2007). These kicks originate from anisotropic
gravitational-wave emission during the final merger. Single BHs that are formed in
such mergers acquires velocities that are presented in Fig. \ref{f.vrec}. 
We find that the velocity distribution is very wide $\sim 10-1000 \kms$, and
that about 70\% of BHs obtain velocities high enough to leave the cluster. It is 
important to note that the shape of the high end of the recoil velocity distribution 
is not well determined. Such strong kicks are obtained by high-eccentricity 
systems which, at present, are not properly modeled (Sopuerta et al.\ 2007). 
Nevertheless, there is no doubt that a majority of BHs obtain kicks higher than 
the cluster ejection velocity ($50 \kms$) and leave the cluster.

\subsection{LIGO detection rates}
\label{ligo}

Fig.~\ref{f.rates} presents the expected range of detection rates of double BH mergers 
for the initial (current stage) and the advanced (expected in 2014) LIGO
detector as a function of cluster mass fraction: $f_{\rm cl}$ -- the ratio of 
stellar mass contained in clusters to the total stellar mass within the reach 
of the detectors. For both stages of the LIGO detector two lines were presented: the upper
limits corresponds to merger rates obtained in this work, the lower limits base
on the O'Leary et al. (2006) results.

Our detection rates were obtained assuming Euclidean space geometry with the
use of inspiral rates for field $R_{\rm MW} \approx 15 {\rm Myr}^{-1}$  calculated 
by Belczynski et al. (2007; for a Milky Way-like galaxy: the sum of averaged 
rates for all double compact mergers in their Model A) and the cluster 
inspiral rate obtained in this work ($R_{\rm cl,0}=2.5 \times 10^{-9} {\rm yr}^{-1}$) 
that scaled up to the stellar mass of the Milky Way is: $R_{\rm cl}=181.5 
{\rm Myr}^{-1}$. We assumed that the star formation rate in the Galaxy was constant
throughout the last 10 Gyr at the rate of $3.5 \mpy$, which results in a total stellar mass
in the Galactic disk of $M_{\rm MW}=3.5 \times
10^{10} \msun$.
The detection rates are calculated using the following formula:
$$
{\cal R_{\rm LIGO}} = \rho_{\rm gal} {4 \pi \over 3} d_0^3 \left((1-f_{\rm cl}){\cal M}_{\rm dis,MW} 
{\cal R_{\rm MW}}+f_{\rm cl}{\cal M}_{\rm dis,cl} {\cal R_{\rm cl}}\right)
$$
where $f_{\rm cl}$ is the cluster mass fraction and $\rho_{\rm gal}=0.01$ Mpc$^{-3}$ 
is the number density of Milky-Way-type galaxies that approximates the mass distribution 
within the LIGO distance range $d=d_0 ({{\cal M}_{\rm c,bhbh} / {\cal M}_{\rm
c,nsns}})^{5/6}$, with $d_0=18.4,\ 300$ Mpc for the initial and advanced LIGO respectively. 
The distance range estimates were obtained for a binary with two
$1.4 \msun$ neutron stars with chirp mass of ${\cal M}_{\rm c,nsns}=1.2 \msun$,
and we rescale them for our populations of double BH binaries for given 
chirp masses ${\cal M}_{\rm c,bhbh}$.  The scaling factor for the cluster population is obtained from
$$ {\cal M}_{\rm dis,cl} = \left< \left( {{\cal M}_{\rm c,bhbh} / {\cal M}_{\rm c,nsns}} 
\right)^{5/2} \right>=1275.0$$
where the average is taken over all BH-BH mergers we got in the simulation. The 
adequate factor for the field population is taken from Belczynski et al.
(2007a; ${\cal M}_{\rm dis,MW} = 71.1$). The specific values for
$\rho_{\rm gal}$, $d_0$ and ${\cal M}_{\rm c,nsns}$ were adopted from
O'Shaughnessy et al.\ (2006). Such a procedure is justified
instead of a more detailed treatment like the one in Bulik, Belczynski \& Rudak (2004),
as there is no significant correlation between the chirp mass
and the merger time of binaries.

The merger rates obtained in the O'Leary et al. (2006) work strongly depend on the stage
of the cluster evolution. Therefore, to calculate the detection rates, they had to use a 
cosmological model. It was described in detail in \S4.1 of their work. The authors assumed 
uniform cluster number density in the universe $\rho_0=1 {\rm Mpc^{-3}}$, which 
corresponds to cluster mass fraction $p=0.0014$ (assuming that all clusters resemble our 
standard model cluster). For the initial stage of LIGO ($d_0=18.4 {\rm Mpc}$) they obtained 
${\cal R}\sim 0.006 {\rm yr^{-1}}$, while for the advanced version of the detector ($d_0=300 {\rm Mpc}$) 
they got ${\cal R}\sim 8 {\rm yr^{-1}}$. We combined their results with the merger rates estimations
for the field population. The total detection rates as a function of the initial cluster mass 
fraction, basing on the assumptions and results of the O'Leary et al. (2006) project, are presented
in Fig.~\ref{f.rates} (the lower limits for both detectors).

The LIGO detection rates presented in Fig. \ref{f.rates} are linearly
increasing functions of the cluster mass fraction $f_{\rm cl}$. Assuming no 
mass in clusters at all ($f_{\rm cl}=0$) we obtain detection rates as calculated 
in Belczynski et al.\ (2007): $0.005 yr^{-1}$ and $19 yr^{-1}$ for LIGO and 
advanced LIGO, respectively. Once cluster contribution is included the increase
of the detection rates is dramatic. The cluster mass fraction is not well known 
but is expected at the level $f_{\rm cl} \gtrsim 0.001-0.01$. Therefore, the
detection rate of merging compact binaries is found to be 
${\cal R_{\rm LIGO}} \gtrsim 0.01-1$ yr$^{-1}$ and $\gtrsim 25-3000$ yr$^{-1}$
for initial and advanced LIGO, respectively.

Dynamical interactions in the cluster core can result in ejection of a
close BH-BH binaries. Most of such events are caused by binary-binary
encounters. The ejected systems gain spatial velocity of the order
$\sim 100\kms$ and continue evolution in an isolation (i.e., evolution
is driven only by emission of gravitational radiation that eventually
leads to a merger). Our study shows that $\sim 10\%$ of BH-BH mergers
originating from clusters take place outside cluster boundaries. The
coalescence takes $\sim 1$ Gyr, the time in which a given system can
travel $\sim 10-100$ kpc and merge in the intergalactic medium.
Therefore, we expect that a small but a significant number of advanced LIGO 
detections of cluster BH-BH mergers will be found far away from host galaxies: 
${\cal R} \sim 30 yr^{-1}$ and $\sim 300 yr^{-1}$ for $f_{cl}=0.001$ and $0.01$, 
respectively.

\subsection{Comparison with other studies}

A number of papers estimating the LIGO detection rates have been published recently. They were
based on various methods and assumptions. To estimate the range of total merger rates we use
the most recent work of O'Leary et al.\ (2006). They used 
an approach similar to ours for cluster dynamics (the encounter rate technique).
They assumed a uniform cluster number density in the universe $\rho_0=1 Mpc^{-3}$, which 
corresponds to a cluster mass fraction $p=0.0014$. Their models did not follow stellar evolution so 
that they had to make several  assumptions which differ from this work:
{\em (i)} BHs were assumed to be completely decoupled from the background cluster stars.
{\em (ii)} All BH-MS binaries at 11 Myr after the cluster formation were assumed to 
eventually become BH-BH systems. 
{\em (iii)} All BHs in O'Leary et al. (2006) were introduced initially into the cluster 
core (they were immediately subject to dynamical interactions).
These different assumptions are likely responsible for the qualitatively different results of 
O'Leary et al.\ (2006): no BH-MS binaries, rapid BH subcluster evaporation, and a predicted merger rate 
rapidly decreasing with time. The detection rates O'Leary et al.\ (2006) obtained (their Fig.\ 8) 
were used as the lower limits for our estimations of the total merger detection rates.

\section{DISCUSSION}
\label{summary}

We have presented the ranges of total cosmic merger rate of double compact
objects in the local Universe. Our results apply to the distances that will
be reached by ground based gravitational-wave detectors such as LIGO or
VIRGO ($\sim 300$ Mpc for a typical NS-NS system). In calculating the 
rates we have considered formation of close double compact objects in 
field populations as well as in dense stellar clusters, i.e., globular 
clusters. The main result of our study shows that the predicted merger rates 
are too small for detection with the current instruments (i.e., initial LIGO) 
but are very promising for the upgraded detectors (i.e., advanced LIGO).  

We find that field populations dominate the formation of close NS-NS and BH-NS
systems. This was already predicted by Phinney (1991) on the basis of observed
binary pulsars. Recently Grindlay, 
Portegies Zwart \& McMillan (2006) have shown that globular clusters can 
contribute only up to $\sim 20 \%$ to the cosmic NS-NS merger rates. 
Formation of double compact object systems with NSs is inhibited by the rather
large natal kicks NSs receive in supernovae (Hobbs et al.\ 2005). These
lead to {\em (i)} disruption of binaries hosting NS progenitors and
{\em (ii)} ejection of NSs from the cluster. The remaining systems containing a NS
interact with heavier stars, and in particular BHs, and through exchange
interactions are removed from binaries and do not form double compact
objects. Also, the initial cluster mass that we use in our simulations
($\sim 5 \times 10^5\msun$, which corresponds to the current average globular cluster mass 
in our Galaxy; e.g., Meylan \& Heggie 1997, see their Fig.~10.2) is too small
to produce a significant number of NS-NS or BH-NS systems. In particular,
in our cluster simulations, we do not find any mergers of NS-NS nor BH-NS
systems.
The results for field populations of NS-NS and BH-NS mergers were
discussed in detail by Belczynski et al.\ (2007a). They find (their model A) 
that field  Galactic merger rates are $\sim 15$ Myr$^{-1}$ and 
$\sim 0.1$ Myr$^{-1}$ for NS-NS and BH-NS, respectively. In our work we
have adopted these rates as the total NS-NS and BH-NS merger rates (field + clusters) 
since the contribution from clusters is rather small for these systems. 
As discussed by Belczynski et al. (2007a) these rates are too small for any
detection with the initial LIGO, while for advanced LIGO a
small but significant number of detections is predicted: $\sim 15$
yr$^{-1}$ and $\sim 1$ yr$^{-1}$ for  NS-NS and BH-NS, respectively. 

Formation of close BH-BH systems that merge in a Hubble time is expected to be very 
effective in clusters. This was already noted in studies that use 
realistic initial conditions for the evolution of BH-BH binaries in clusters (e.g., 
Miller \& Hamilton 2002; Gultekin, Miller \& Hamilton 2004; O'Leary et 
al.\ 2006). Here, we have followed the evolution of a
realistic average cluster with full stellar evolution and physical 
treatment of all BH-BH binaries, assuming that binary interactions are able
to prevent the BHs from separating into an isolated subcluster. 
We have found that the production of BH-BH binary
mergers under these assumptions is indeed remarkably effective in dense clusters.
The most striking and counter-intuitive difference with other 
studies is that the predicted BH-BH merger rate in clusters is rather 
constant over long periods of time ($\sim$ Hubble time). In earlier
studies the BH populations were usually introduced in the cluster core and
evolved separately from the other stars in the cluster. That leads to a
larger merger rate at early times in the cluster evolution followed by a very  rapid 
drop in the merger rates once BHs are eliminated (through mergers and ejections).  
However, the assumptions introduced in our work, namely {\em (i)} stars
that form BHs are initially placed throughout the cluster, 
and {\em (ii)} BHs do not segregate so strongly as to form a completely decoupled
subcluster and are instead allowed to interact 
with other stars in the cluster, make our results qualitatively different. We find
that BHs that form in the core, in fact, produce mergers at the early stages
of cluster evolution. But later many massive stars that were in the halo and
that formed BHs  will steadily feed the cluster core with BHs, i.e., 
continued mass segregation in the cluster halo is providing BHs to the cluster core on 
long timescales. 
Additionally, exchange interactions of BHs with unevolved (e.g., main 
sequence stars) in the core are effective over a long periods of time, and
usually lead to formation of close BH-BH systems that merge later in the
core. A simplified scenario involves two single BHs sinking into the core;
each catches a main sequence companion, and then two BH-MS binaries
interact, forming a close BH-BH system and two single main sequence stars are 
released back into the cluster. 

\begin{figure}
\epsscale{.55}
\plotone{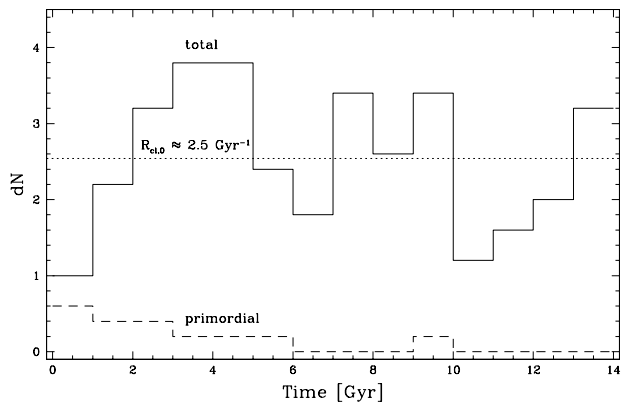}
\caption{
Number of BH-BH mergers versus time averaged over 5 independent runs
($M_{cluster}\approx4.82\times 10^5 \msun$). The dashed line shows the number of BH mergers 
coming from primordial BH binaries (the bottom-most channel in Fig.\ref{f.channels}).
Note that all numbers were binned per Gyr, so the rate of BH-BH mergers ($Gyr^{-1}$) can be read
directly off this plot and the average (approximately constant) merger rate
is indicated by the dotted line.
}
\label{f.merg}
\end{figure}

Our average cluster merger rate for BH-BH systems is $\sim 3$ Gyr$^{-1}$
(see Figure~\ref{f.merg}) for our simulated cluster with mass 
$M_{\rm cl}=4.8 \times 10^5 \msun$. If we scaled this rate to the mass of the
galactic disk ($M_{\rm MW}=3.5 \times 10^{10} \msun$), then the BH-BH merger
rate would come up to $\sim 180$ Myr$^{-1}$ (i.e., this is the expected BH-BH 
merger rate if the entire mass of our Galactic disk were contained in dense globular
clusters). We should compare this rate to the Galactic field BH-BH merger
rate, $0.025$ Myr$^{-1}$ (Belczynski et al.\ 2007a). It is clear
that the production of BH-BH mergers is much more efficient (by $\sim 3-4$ 
orders of magnitude) in clusters as compared to field evolution\footnote{
We have used the rate from the same evolutionary model used for field
population (model A of Belczynski et al.\ 2007a) as it was employed in
our standard cluster model.}. 

To combine our predicted cluster rates with the field rates we need to know the initial
stellar mass fraction contained in clusters. If we look at the Galactic globular
clusters we find that they contain about 0.001 of the total mass in stars 
found in the field ($f_{\rm cl}=0.001$). However, it is reasonable
to expect that many clusters may have been completely destroyed and that
the clusters we see today were initially more massive and lost significant
mass through evaporation (e.g., Vesperini 1998; Joshi et al.\ 2001). Although we present
 our predictions for the entire
range of plausible $f_{\rm cl}$ values, we consider that the most reasonable estimate
is still $f_{\rm cl} \gtrsim 0.001$. While we do not have 
reliable mass estimates for globular clusters in elliptical 
galaxies, it has been shown that the specific frequency of GCs per galaxy luminosity
in ellipticals is significantly (about an order of magnitude) higher than in spiral galaxies 
(Kim \& Fabbiano 2004). Moreover, elliptical galaxies are on average 
more massive than spiral galaxies. Therefore, an upper limit on the initial mass fraction
contained in all clusters, although highly uncertain, can probably be set to $f_{\rm cl} \lesssim 0.01$.

The total cosmic merger rate for BH-BH systems is then strongly dependent on
the mass contained in globular clusters. For a small fraction
($f_{\rm cl}=0.001$) we find the merger rate per Milky Way in our
model to be $\sim 0.2$ Myr$^{-1}$, 
while for a larger fraction ($f_{\rm cl}=0.01$) the merger rate is $\sim 2$ Myr$^{-1}$.
At this point we can also estimate the detection rate for a given GW detector. We
must keep in mind that the average chirp mass of field BH-BH binaries is much
smaller ($M_{\rm c} \sim 7 \msun$) than for cluster BH-BH mergers ($M_{\rm
c} \sim 20 \msun$). This allows us to observe cluster BH-BH mergers in a much
larger volume and their relative contribution to the detection rate is
greater than indicated simply by the merger rates (see \S\,~\ref{ligo}). 
One could also imagine forming of few very massive BHs (up to $\sim 100\msun$) in the 
cluster (e.g., through mergers of massive binaries; Belczynski et al.\ 2006) which 
could form BH-BH mergers characterized by extremely high chirp masses. 
Such mergers would be detectable from much greater distances, making the observed 
rates even higher. 

The predicted ranges of total detection rates, for all types of double compact objects, are 
presented in Figure~\ref{f.rates} as a function of cluster contribution. The most 
likely range of values for this parameter ($f_{\rm cl} \sim 0.001\div 0.01$) is marked with
the vertical shaded area in Figure~\ref{f.rates}.
For very low cluster contributions ($f_{\rm cl}=0.0001$) the detection rates
correspond to mergers coming only from field populations and are adopted
from the reference model of Belczynski et al.\ (2007b; their model A). With
increasing cluster contribution we see a drastic increase in predicted detection
rates. This increase is connected to the very effective production of BH-BH
mergers in clusters as discussed above. For advanced LIGO the detection 
rates could be as high as $\sim 25-3000$
yr$^{-1}$ and are higher by more than an order of magnitude than the rates 
just for field populations. The total rates are dominated by
dynamically formed BH-BH mergers in dense stellar clusters. 
If advanced LIGO does not observe this population of BH-BH mergers it
will put strong constraints on the initial stellar mass fraction contained in
dense stellar clusters.   

\begin{figure}
\plotone{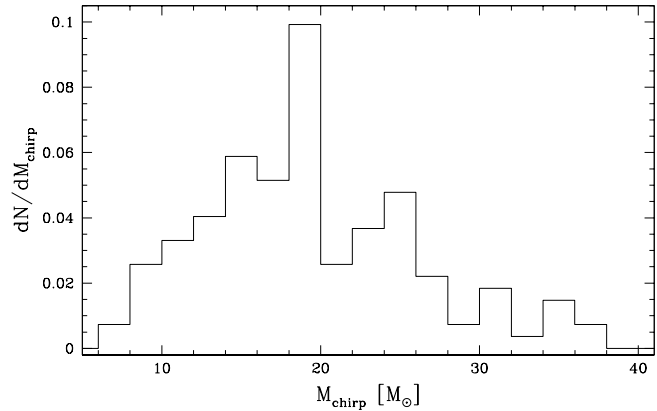}
\epsscale{.55}
\caption{
Distribution of chirp masses ($(M_1M_2)^{3/5} (M_1+M_2)^{-1/5}$) of merging BH-BH binaries in 
the cluster population.
}
\label{f.mchirp}
\end{figure}

\begin{figure}
\plotone{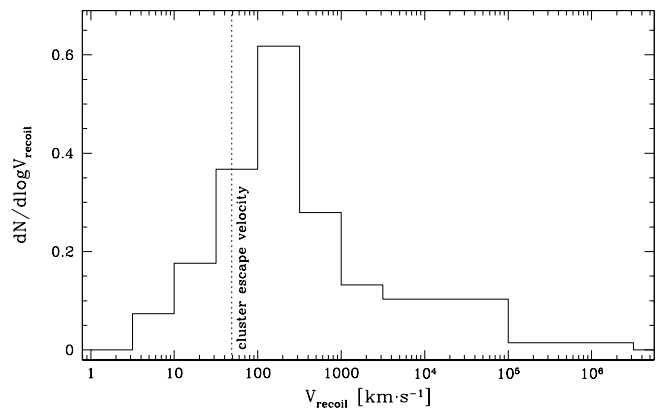}
\epsscale{.55}
\caption{
Distribution of recoil velocities obtained by BH-BH mergers due to anisotropic GW emission. 
About 70\% of BHs get kick velocities higher than the cluster escape speed. Note the 
horizontal axis logarithmic scale.
}
\label{f.vrec}
\end{figure}

\begin{figure}
\epsscale{.75}
\plotone{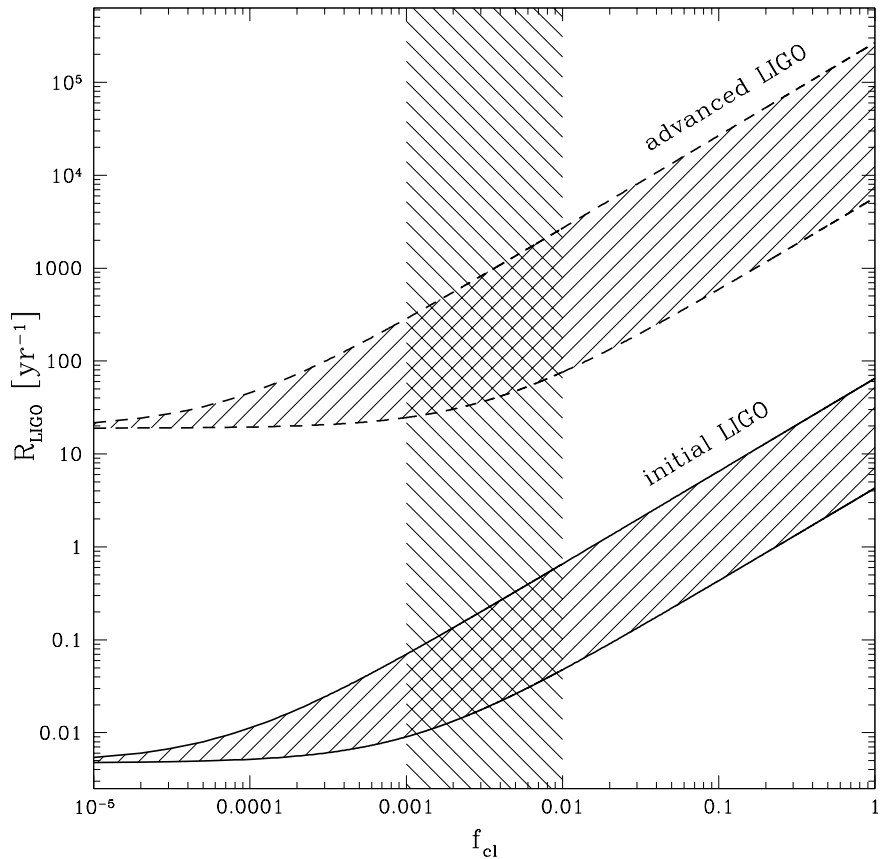}
\caption{
LIGO and Advanced LIGO detection rates of merging double compact binaries as a function of the
assumed initial stellar mass 
fraction in dense star clusters. The vertical shaded area represents the most likely range of this parameter (see
\S\,4 for more details). For both versions of the LIGO detector two lines are shown: the lower limit
is based on the cluster merger rates obtained by O'Leary et al.\ (2006) while the upper limit is based 
on the model described in this work.
}
\label{f.rates}
\end{figure}

The production of BH-BH mergers in the field is inhibited by the process
identified in Belczynski et al.\ (2007b): many potential BH-BH progenitors
evolve through a common envelope phase while the donor is evolving through
the Hertzsprung gap. Such a common envelope leads most likely to a merger
and aborts potential formation of a BH-BH system (because, in the
Hertzsprung gap, the star has not yet developed a clear core-envelope structure and
the inspiral does not stop before complete merger of the two interacting stars). 
If this current understanding of the common envelope phase is correct, we do not expect
detection of more than a few field BH-BH mergers per year. However, if the
progenitors somehow survive this phase, we could then expect up to $\sim100$ detections of
field BH-BH mergers per year by Advanced LIGO. 
Since the chirp mass distribution of
the field and cluster populations are so different (see Figure~\ref{f.mchirp}
and Figure 4 of Belczynski et al.\ 2007b) it would be easy to tell the two
populations apart. This would, in turn, allow us to both {\em (i)} derive the 
initial mass fraction in clusters, and {\em (ii)} constrain the fate of massive 
binary systems going through a common envelope phase. This result highlights the importance 
of the chirp mass distribution as a diagnostic tool in gravitational wave astronomy
(Bulik \& Belczynski, 2003; Bulik, Belczynski, \& Rudak, 2004).

\acknowledgements
AS, KB and TB acknowledge support from KBN grants 1P03D02228 and 1P03D00530.   
FAR acknowledges support from NSF Grant PHY-0601995 and NASA Grant NNG06GI62G
to Northwestern University. We also thank anonymous referee for practical report that helped us in presentation of our results.

\clearpage

\acknowledgements

\end{document}